\documentclass[aps,prb,longbibliography,floatfix,superscriptaddress,showpacs,amsmath,amssymb,reprint,nofootinbib]{revtex4-2}

\usepackage{graphicx}
\usepackage[english]{babel}
\usepackage{bm}
\usepackage{bbm}
\usepackage{amsmath,amsfonts}
\usepackage{amsbsy}
\usepackage[colorlinks=true,linkcolor=blue,urlcolor=blue,citecolor=blue,breaklinks=true]{hyperref}
\usepackage[utf8]{inputenc}

\usepackage{graphics}
\usepackage{subcaption}
\usepackage{xcolor}
\usepackage{bbm}
\usepackage{graphics}
\usepackage{graphicx}
\usepackage{epsfig}
\usepackage{amsmath}
\usepackage{amssymb}
\usepackage{amsfonts}
\usepackage{bm}
\usepackage{color}
\usepackage{xcolor}
\usepackage{bbm}
\usepackage{hyperref}
\usepackage[normalem]{ulem}
\newcommand{\addAA}[1]{{\color{blue}{#1}}}

\begin{document}

\title{ Properties of two level systems in current-carrying  superconductors}

\author{T. Liu}
\affiliation{Department of Physics, University of Wisconsin-Madison, Madison, Wisconsin 53706, USA}
\author{A.~V. Andreev}
\author{B.~Z. Spivak}
\affiliation{Department of Physics, University of Washington, Seattle, WA 98195,  USA}

\date{\today}

\begin{abstract}
We show that in disordered $s$-wave superconductors, at sufficiently low frequencies  $\omega$, the coupling of two level systems (TLS) to external \emph{ac} electric fields increases dramatically
in the presence of a \emph{dc} supercurrent. This giant enhancement manifests in all \emph{ac} linear and nonlinear phenomena. In particular, it leads to a parametric enhancement of the real part of the \emph{ac} conductivity and, consequently, of the equilibrium current fluctuations. If the distribution of TLS relaxation times is broad, the conductivity is inversely proportional to $\omega$,  and the spectrum of the equilibrium current fluctuations takes the form of $1/\mathrm{f}$ noise.
\end{abstract}

\maketitle

\section{Introduction}

\begin{figure}[t]
\centering
\includegraphics[width=.8\linewidth]{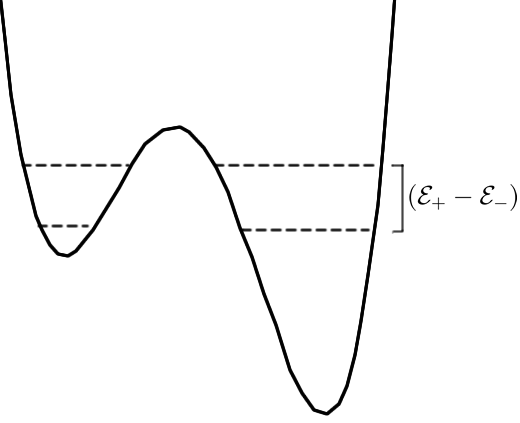}
\caption{Qualitative picture of a TLS potential plotted as a function of the configurational coordinate. The dashed lines represent the energy eigenvalues in the two states of the TLS, which differ by $\left(  \mathcal{E}_+ - \mathcal{E}_-  \right)$ .
}
\label{Fig1}
\end{figure}

The concept of two-level systems (TLS) was introduced to explain the unusual thermal and transport properties of insulating and metallic glasses~\cite{AGV, Phyl} (see Refs.~\cite{Phillipss, Black, GGK, Muller, Hunklinger, Burin} for a review of the subject).
While the microscopic nature of TLS is still not  fully understood, the  standard phenomenological  model
assumes the ability for an atom, or several atoms, to tunnel quantum mechanically at low temperatures between two quasi-stable atomic configurations, as illustrated in Fig.~\ref{Fig1}.

The dynamics of TLS  may be described by a $2\times2$ matrix Hamiltonian of the form
\begin{eqnarray}\label{H}
\hat{H}_0= \begin{pmatrix}
\epsilon & t \\
t^* & - \epsilon  \\
\end{pmatrix}.
\end{eqnarray}
Here $2\epsilon$ is the energy difference between the two quasi-stable configurations and $t$ is the tunneling amplitude between them.
Diagonalization of this Hamiltonian  yields the eigenvalues of the system
  \begin{equation}\label{E}
  \mathcal{E}_{\pm} =\pm \sqrt{\epsilon^{2}+|t|^{2}}.
\end{equation}
External perturbations modify the parameters of the Hamiltonian. Usually, it is assumed that the modulation of the inter-well splitting $\epsilon$ is much bigger than the modulation of $t$  (see for example Ref~\cite{Hunklinger}). Thus, the coupling of TLS to external low frequency perturbations is described by the time-dependence of the inter-well energy splitting $\epsilon(t) = \epsilon_0 + \delta \epsilon (t)$. For example,  in the presence of an external electric field ${\bf E}(t)$ we have a ``dipole" contribution to the energy splitting,
\begin{align}\label{eq:epsilon_d}
\delta \epsilon_d (t) = & \,   - {\bf E}(t)\cdot {\bf d},
\end{align}
where $2{\bf d}$ is the dipole moment difference between the two metastable  states of the TLS
whose magnitude and direction are random.   The coupling of this form changes both the  eigenvalues $ \mathcal{E}_{\pm}$ and the eigenfunctions, and therefore adequately describes both resonant and relaxation absorption mechanisms of the electric field by the TLS.

The TLS has also been suggested as a source of the ubiquitous $1/\mathrm{f}$ noise in conductors  (see for example \cite{Data, Kogan, Weismann, GalperinAltshuller}).
Since the two states of the TLS have different electron scattering cross-sections, transitions between them change the local conductivity. If  the distribution function of TLS relaxation times is broad, the
$1/\mathrm{f}$ noise of current fluctuations through a voltage-biased sample can be attributed to fluctuations in sample conductance induced by the TLS transitions. This picture has been confirmed by measuring  ``noise of the noise" in metals \cite{clarke}.

At low temperatures $T$, the quantum interference of electron waves in metals results in
large sensitivity of electron transport to the motion of individual scatterers  \cite{AltSpivSenc, FengLeeSen}.
At relatively weak magnetic fields,  the electron system crosses over from the ``orthogonal'' to the ``unitary'' ensemble and, as a result,  the amplitude of the $1/\mathrm{f}$ noise is suppressed by a factor two ~\cite{FengLeeSen, Birge1/fH}.
Conversely, due to the interaction of TLS with the Friedel oscillations of the electron density,
the properties of TLS in metals at low temperatures are strongly affected by the quantum interference of the electron  wave functions, resulting in a strong  dependence of the TLS activation energy  on the magnetic field
\cite{AltshullerSpivakH}.
This phenomenon was revealed by measurements of the magnetic field dependence of the plateaus in the telegraph signal associated with transitions between the two states of a single fluctuator  \cite{BirgeH, NormanTelNoise}.
Thus, the strong interaction of TLS with the local density of electrons in disordered conductors has been verified experimentally.

In disordered $s$-wave superconductors, the TLS become the most abundant low-energy excitations at low temperatures and thus play important role in the dissipation and decoherence of superconducting systems. Since the typical  size of  TLS is believed to be smaller than the superconducting coherence length  $\xi $, their atomic structure and strength of coupling to the electron density should be similar to that in normal metals.

In this article, we consider the effect of superconducting pairing  on the coupling of TLS to the external ac electric field ${\bf E}(t)= \mathrm{Re} \large( {\bf E}_\omega e^{-i\omega t} \large)$.
We show that in the presence of a \emph{dc} supercurrent and at sufficiently small frequencies, the effective
interaction between the TLS and the electromagnetic field is parametrically enhanced.  The physical origin of this  effect lies in the large sensitivity of the random Friedel oscillations of the electron density $n({\bf r}, {\bf p}_{s})$ in disordered superconductors to the changes of the superfluid momentum
\begin{equation}\label{eq:p_s}
{\bf p}_{s}(t)=\frac{\hbar}{2} \bm{\nabla} \chi-\frac{e}{c}{\bf A}(t).
\end{equation}
Here $\chi ({\bf r},t)$ is the phase of the superconductiviting order parameter, and ${\bf A}(t)$ is the vector potential.
The inter-well energy splitting $\epsilon (t)$ of the TLS  is affected by the interaction of the atoms forming TLS with the electron density.  Therefore  it  acquires a ${\bf p}_{s}$-dependent correction $\delta \epsilon_s $.  Since atomic TLS do not break time reversal invariance, their energy must be an even function of ${\bf p}_s$.
Therefore, at small supercurrent densities this correction can be written in the form
\begin{align}\label{eq:epsilon_p_s}
\delta \epsilon_s(t) = & \,  \alpha_s\,    p_s^2 (t) , \quad \alpha_s \equiv \left. \frac{d \epsilon (p_s^2)}{ d \left( p_s^2\right) }\right|_{p_s^2 =0} .
\end{align}
In the presence of an \emph{ac} electric field  the superfluid momentum  becomes time-dependent,
\begin{equation}\label{eq:Jos}
\dot{{\bf p}}_{s}(t)=e{\bf E}(t).
\end{equation}
This causes time-dependence of the TLS Hamiltonian in  Eq.~\eqref{H}.   Equation \eqref{eq:epsilon_p_s} shows that a linear coupling of the external electric field to TLS mediated by superconductive pairing is possible only in the presence of a \emph{dc} superfluid momentum $\bar{\bf p}_s$.
In this case, the time-dependent
superfluid momentum can be written as ${\bf p}_{s}(t)=\bar{{\bf p}}_{s} +\delta {\bf p}_{s}(t)$,  where, according to
Eq.~\eqref{eq:Jos},  $\delta {\bf p}_{s}(t)=\mathrm{Re}\large(i e{\bf E}_{\omega} e^{-i\omega t}/\omega\large)$.

The instantaneous relation between the time-dependent superfluid momentum ${\bf p}_s(t)$ and the energy splitting described by Eq.~\eqref{eq:epsilon_p_s} remains valid as long as $\omega \ll \Delta$.
Combining Eq.~\eqref{eq:epsilon_d} with Eq.~\eqref{eq:epsilon_p_s}, linearized with repect to $\delta {\bf p}_{s}$, we find that  in this frequency interval, the interaction of TLS with an \emph{ac} electric field   in  current-carrying superconductors has the form $\delta \epsilon(t) = \Re \left( \delta \epsilon_\omega e^{- i \omega t}\right)$ where
\begin{align}\label{eq:delta_epsilon_omega}
\delta \epsilon_\omega =  & \, - \left( \bm{d}   - i \frac{e}{\omega } \alpha_s \bar{\bf p}_s \right) \cdot {\bf E}_\omega .
\end{align}
Although Eqs.~\eqref{eq:epsilon_p_s}  and \eqref{eq:delta_epsilon_omega} can be introduced phenomenologically, evaluation of the parameter  $\alpha_s$, which characterizes the 
sensitivity of TLS level splitting to the superfluid momentum, requires a microscopic consideration.

\begin{figure}[t]
\centering
\includegraphics[width=.9\linewidth]{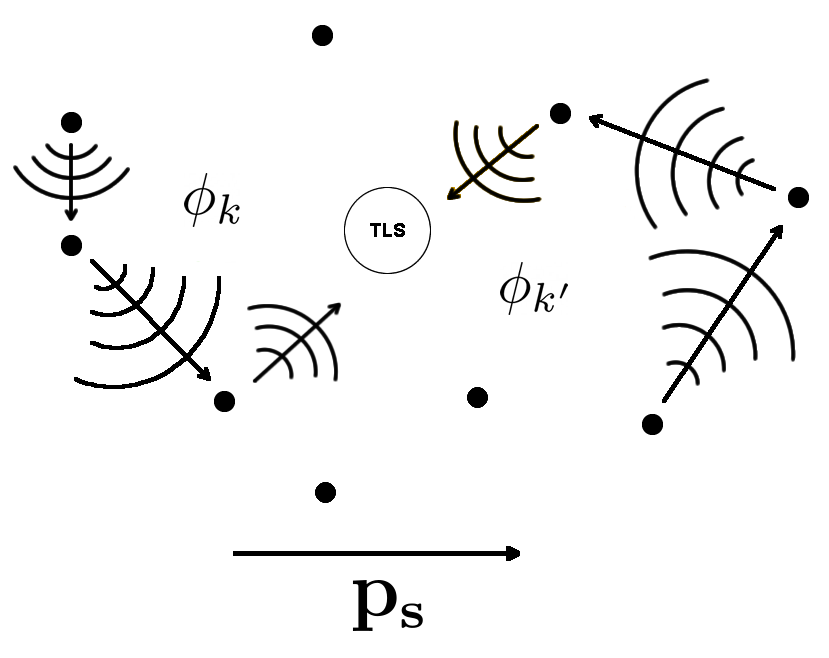}
\caption{  Illustration of the  Friedel oscillations caused by interference of electron waves scattered from multiple impurities.
The black dots represent randomly distributed impurities.
}
\label{Fig1.jpeg}
\end{figure}

Below, we develop a microscopic theory of this effect for \emph{s}-wave superconductors in the diffusive regime,  where the coherence length $\xi$   is larger than the electron mean free path $\ell$ in the normal state, $\xi = \sqrt{D/\Delta}>\ell>\lambda_{F}$. Here   $\Delta$ is the superconducting gap, $D=v_{F}\ell /3$ is the diffusion coefficient,  $v_{F}$ is the Fermi velocity, and $\lambda_{F} $ is the Fermi wavelength. In this case, the magnitude of the parameter  $\alpha_s$ is large because of the high sensitivity of the Friedel oscillations of the electron density to the changes in $p_s$.
We assume that the typical  kinetic and potential energy of electrons in the system are of the same order,  $r_{s}=e^{2}/\hbar v_{F} \sim 1$, and that
the characteristic size of TLS is of the order of the interatomic spacing, which in metals is of the order of $\lambda_{F}$.
In this case,   the $p_s$-dependent correction to inter-well energy splitting $ \delta \epsilon_{s}$ in Eq.~\eqref{eq:epsilon_p_s} may be estimated as
\begin{equation}\label{eq:epsilon_s_estimate}
\delta \epsilon_{s} (t) \sim e\lambda_{F} \delta n({\bf r}, {\bf p}_{s} (t)),
\end{equation}
where $\delta n({\bf r}, {\bf p}_{s} (t))$ is the $p_s$-dependent correction to the electron density.   Thus, the
parameter $\alpha_s$ in Eqs.~\eqref{eq:epsilon_p_s} and \eqref{eq:delta_epsilon_omega} can be expressed in terms of the sensitivity of the variations of the electron density $\delta n({\bf r}, {\bf p}_{s} (t))$ to the superfluid momentum. This enables us to evaluate the coupling of TLS to \emph{ac} electric fields and all subsequent quantities  within a factor of order unity.

\begin{figure}[t]
\centering
\includegraphics[width=.9\linewidth]{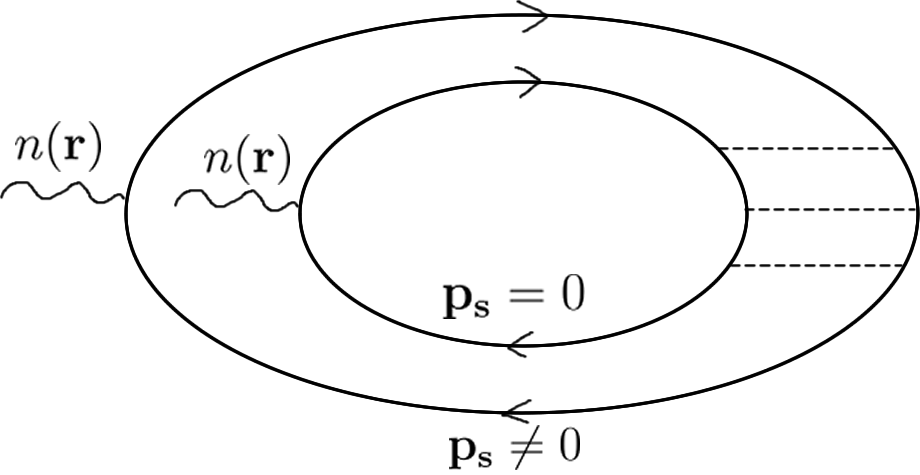}
\caption{
Feynman diagrams for calculating $\langle (\delta n)^{2} \rangle$. The dashed lines represent disorder averaging.
The solid line correspond to the electron  Green's functions, which are matrices in the Gorkov-Nambu space. }
\label{Fig2}
\end{figure}

For simplicity, we consider the situation where a superconducting film with a thickness $L_{z}$ smaller than the skin length is exposed to a monochromatic \emph{ac} electric field of frequency $\omega$. The spatial modulations of the electron density $ n({\bf r})$, which contribute to the TLS level splitting are caused by the interference of the electronic waves scattered by different impurities.
Due to the rapid decay of the Friedel oscillation amplitude with the distance from the impurity,  the value of  $n({\bf r})$  is determined primarily by impurities closest to the TLS.  However, the sensitivity of the electron density to the changes in ${\bf p}_s$ is determined by the interference of quantum  amplitudes for the electron propagation from the impurities separated from the observation point by distances of order $\xi$. This can be understood as follows:
At ${\bf p}_{s}\neq 0$, the electron amplitudes corresponding to different diffusion paths acquire random phases $\delta \phi_{k}\sim \int {\bf p}_{s} \cdot d {\bf l}_{k}$, where
 the index $k$ labels different diffusion paths.
This changes the interference of contributions of different paths to the electron density.
Since $\delta \phi_{k}$ increases with the length of the path,   the sensitivity of the electron density $\delta n ({\bf r},  {\bf p}_{s})=n({\bf r}, 0)-n({\bf r}, {\bf p}_{s})$ is controlled by contributions from electron waves traveling from impurities situated at distances of the order of the superconducting correlation length  $\xi $ from point ${\bf r}$.
This is shown qualitatively in Fig.~\ref{Fig1.jpeg}.
Therefore,  the variance of $\delta n ({\bf r},  {\bf p}_{s})$ in the diffusive regime can be obtained using the standard diagram technique of averaging over random impurity potential \cite{Abricosov}, valid at $\lambda_{F}\ll \ell$. To be concrete, we consider thin films of thickness smaller than the coherence length, $\lambda_{F}<L_{z}<\xi$.
Evaluating the Feynman  diagrams shown in Fig.~\ref{Fig2}, we get
\begin{eqnarray}\label{eq:deltan}
\sqrt{\langle (\delta n)^{2} ({\bf r},  {\bf p}_{s}) \rangle} =
\frac{C}{\lambda_{F}^{3}\sqrt{G}}\frac{\Delta}{\epsilon_{F}}\left( \frac{p_{s}}{p_{d}}\right)^{2}, \quad  p_{s}\ll p_{d}.
\end{eqnarray}
Here $\langle \ldots \rangle$ denotes averaging over disorder,  $\epsilon_{F} $ is the Fermi energy,  $p_{d}=\sqrt{\Delta/D}$  is the depairing superfluid momentum, $C $ is a  numerical factor of order unity, and
\begin{equation}
G= \frac{8 \pi}{3} \frac{\ell L_{z}}{\lambda_{F}^{2}}
\end{equation}
is the conductance of the film in units of $e^{2}/ 2\pi \hbar$.

Using Eq.~\eqref{eq:deltan} we  obtain the following estimates  for the variances of   $\delta \epsilon_s (t)$ in Eq.~\eqref{eq:epsilon_s_estimate},  and  the parameter $\alpha_s$ in Eqs.~\eqref{eq:epsilon_p_s} and \eqref{eq:delta_epsilon_omega},
\begin{subequations}\label{eq:delta_epsilon_alpha_s_estimate}
  \begin{align}
  \label{eq:deltaepsilon}
  \sqrt{\langle (\delta \epsilon_{s})^{2} \rangle} \sim & \,   |\Delta| \frac{1}{\sqrt{G} } \left(\frac{p_{s}}{p_{d}}\right)^{2} =  \frac{1}{\sqrt{G} } Dp_{s}^{2}, \\
  \label{eq:alpha_s_variance}
  \sqrt{\langle \alpha_s^2 \rangle} \sim & \, \frac{D}{\sqrt{G}}, \quad  \langle \alpha_{s} \rangle =0.
\end{align}
\end{subequations}
These results show that in situations where the electric field is not orthogonal to
$ \bar{{\bf p}}_{s}$,  the  ratio of the second and the first terms in the brackets in Eq.~\eqref{eq:delta_epsilon_omega}
is of order
\begin{equation}\label{omega*}
\frac{e \alpha_{s} \bar{p}_{s}}{\omega  d } \sim  \frac{\omega{*}}{\omega} .
\end{equation}
Here we introduced the characteristic frequency scale defined by the \emph{dc} superfluid momentum
\begin{equation}
\omega^{*}
=\Delta \frac{\bar{p}_{s}}{p_{d}}
\sqrt{\frac{\xi_{0}}{L_{z}}} ,
\end{equation}
with $\xi_{0}=v_{F}/\Delta$ being the coherence length of  pure superconductor.   It follows from Eqs.~\eqref{eq:delta_epsilon_omega}, and \eqref{omega*} that for
\begin{equation}
\omega<  \omega^{*}.
\end{equation}
The interaction between the TLS and the external electric field is dominated by the temporal oscillations of the electron density, rather than by the direct interaction of the external electric field with the dipolar moment of the TLS.

The aforementioned  $\bar{{\bf p}}_{s}$-dependent  enhancement of the coupling of TLS to electric fields manifests itself in all \emph{ac}
phenomena. This includes resonant absorption,  absorption related to the relaxation mechanism, and all nonlinear effects.  The frequency dependences  of these effects will be very different from the conventional ones.

Equations \eqref{eq:epsilon_p_s}, \eqref{eq:delta_epsilon_omega} and \eqref{eq:delta_epsilon_alpha_s_estimate}
show that in the presence of supercurrent, the dissipative part of the conductivity tensor becomes anisotropic.
The enhancement of TLS coupling to \emph{ac} electric fields occurs only for the longitudinal geometry, where the \emph{ac} electric field is parallel to $\bar{\bf p}_s$.   At small $\omega$, the corresponding longitudinal component of the dissipative conductivity, $\sigma_{\|}$,  is significantly enhanced as compared to the   $\bar{\bf p}_{s}=0$ case. In contrast,  the transverse  component $\sigma_{\perp}$
remains roughly the same as that at $\bar{\bf p}_{s}=0$.

As an example, we evaluate the TLS contribution to the dissipative part of the longitudinal microwave conductivity  at sufficiently small frequencies,
where it is controlled by the  Debye relaxation mechanism (see, for example, \cite{Debye, Hunklinger, Shklovskii, Shklovskii1}).
Energy dissipation in the relaxation mechanism is caused by the time-dependence of the TLS energy levels, which creates a non-equilibrium level population, whose relaxation leads to  entropy production.
In this case the main contribution to the absorption power $P=\sigma_{\|} E^{2}_{\omega}/2$ comes from the TLS with  $|\mathcal{E}_+-\mathcal{E}_-|  \lesssim T$,
and   long relaxation times, $ \tau\sim 1/ \omega$. The latter condition corresponds to  $| t|\ll |\epsilon |$ in Eqs.~\eqref{H}, and \eqref{E}, which implies $|\mathcal{E}_+-\mathcal{E}_-| \approx 2 |\epsilon|$. Using  standard arguments (see for example \cite{Shklovskii}),  we  get  the following estimate for the dissipative  part of the  longitudinal  conductivity
\begin{equation}\label{eq: Pdebye}
\sigma_{\|}(\omega) \sim  e^{2} \nu(T)\int d \tau  f(\tau)\frac{ \lambda_{F}^{2}\tau  \left[\omega^{2} + \eta \cdot  ( \omega^{*})^{2}\right] }{1+(\omega \tau)^{2}}.
\end{equation}
Here $\nu (T)$ is the density of states of TLS  with $\epsilon \sim T$  per unit area, $f(\tau)$ is the distribution function of relaxation times normalized to unity,  and $\eta\sim 1$ is a factor of order one.  In general,  $\nu(T)$ is temperature-dependent.  However,  we note that in glasses this quantity does not depend on $T$.

 At  $T\ll \Delta$ the  concentration of quasiparticles in \emph{s}-wave superconductors is exponentially small, and   the  TLS relaxation times are controlled by their interactions with phonons, similar to the case of dielectrics. The frequency dependence of the conductivity depends on how quickly $f(\tau)$ decays at large values of $\tau$.
If   $f(\tau \rightarrow \infty)$ decays quicker than $1/\tau^{2}$ then one can introduce a characteristic   relaxation time $\tilde{\tau}$.
In this case at $p_{s}=0$ the conductivity,  $\sigma(\omega)  \sim e^{2}\nu \lambda_{F}^{2}\omega^{2}\tilde{\tau}$,  is  isotropic and quadratic in frequency.
However, at $\bar{{\bf p}}_{s}\neq 0$ and  $\omega \ll  \omega^{*}$,  the longitudinal component of the conductivity tensor, $\sigma_{\|} (\omega)$, exhibits a giant enhancement,
proportional to the parameter   $(\omega^{*}/\omega)^{2}$.
 In this case
\begin{equation}
\sigma_{\|}(\omega\rightarrow 0)   \addAA{\sim } e^{2}\nu ( \lambda_{F}\omega^{*})^{2}\tilde{\tau}
\end{equation}
 becomes frequency-independent.

In many disordered systems, especially in glasses,  the relaxation time of TLS has an exponential dependence,  $\tau\sim \exp(\zeta)$,  on some parameter $\zeta$,  which is uniformly distributed.  In such cases the distribution function $ f(\tau)$ is broad,  and has the form (see for example  \cite{Kogan,Weismann}),
\begin{equation}
 f(\tau)\sim 1/\tau.
 \end{equation}
Then, according to Eq.~\eqref{eq: Pdebye} at  $\bar{\bf p}_{s}=0$ has a linear dependence on the frequency,
\begin{equation}\label{omega}
\sigma(\omega) \sim e^{2}\nu \lambda_{F}^{2}\omega.
\end{equation}
In contrast, at   $\bar{p}_{s}\neq 0$ , and $\omega\ll \omega^{*}$ the longitudinal conductivity is inversely proportional to $\omega$,
  \begin{equation}\label{sigmaOmega}
 \sigma_{\|}(\omega) \sim e^{2} \nu \frac{( \lambda_{F}\omega^{*})^{2}}{\omega}.
 \end{equation}
It may seem counter-intuitive that according to Eq.~\eqref{eq: Pdebye}, at  $\bar{p}_{s}\neq 0$  the conductivity associated with localized TLS does not vanish as $\omega \rightarrow 0$.
The reason for this is as follows. Inelastic processes induce transitions between the two states of TLS and modulate the supercurrent by changing the superfluid density.   At low frequencies, this modulation turns out to be in phase with the oscillations of the external electric field, leading to dissipation.

Since a superconductor carrying a \emph{dc} supercurrent is in an equilibrium state,  one can use the fluctuation-dissipation theorem (FDT) \cite{Landau}  to relate the spectral density of fluctuations of the total current through the system to the dissipative conductivity
$\sigma_\parallel(\omega)$,
\begin{equation}\label{S}
S_{s}(\omega)=\int dt \overline{ \delta I_{s}(t) \delta I_{s}(0) } e^{i\omega t}= T \sigma_{\|} (\omega), \,\,\,\,\,\,\  \omega <T.
\end{equation}
Here  the subscript  $s$ indicates that the system is in the superconducting state and $\delta I_{s}(t)$ describes
fluctuations component of the total current in the direction of $\bar{{\bf p}}_{s}$.

It follows from Eqs.~\eqref{eq: Pdebye}, and \eqref{S} that in the presence of super-current, and at $\omega< \omega^{*}$   the amplitude of the current fluctuations exhibits giant enhancement.
In particular, according to Eqs.~\eqref{sigmaOmega}, \eqref{S},  in systems  where  $f(\tau)\sim 1/\tau$  the current  correlation function has 1/f form
\begin{equation}\label{1/f_sup}
S_{s} (\omega)\sim \Gamma\frac{I_{s}^{2}}{V\omega} , \,\,\,\,\,\,\,\,\  \Gamma=\frac{\nu T \ell}{\langle N_{s} \rangle ^{2}}.
\end{equation}
Here   $I_{s}$ is the total  bias supercurrent passing through the system, and we used the relation between the average supercurrent density and superfluid density per unit area  $\langle {\bf j}_{s}\rangle =e\langle N_{s}\rangle  {\bf p}_{s}/m$, which is  valid
at $ p_{s}\ll p_{d}$.

The expression for the spectral density of current fluctuations in  Eq.~\eqref{1/f_sup}  describing $1/\mathrm{f}$  noise in superconductors is similar to the corresponding expression for the  $1/\mathrm{f}$ noise of normal current in metals (see for example \cite{Kogan, Weismann}). In both cases the spectral functions are quadratic in the \emph{dc} current and inversely proportional to the volume of the sample.  However,  there are significant differences between these two situations.

 In normal metals, the $1/\mathrm{f}$ noise of the total current is related to the conductance fluctuations and exists only in the presence of a  bias current. These current fluctuations are non-equilibrium and therefore do not obey the fluctuation-dissipation theorem.   In particular, at $\omega\to 0$ the conductivity of metals becomes frequency-independent. In contrast,   Eq.~\eqref{1/f_sup} describes equilibrium current fluctuations, and the divergence of $S_{s} (\omega)$ at $\omega \rightarrow 0$ is accompanied by the corresponding divergence of the conductivity in Eq.~\eqref{sigmaOmega}.

According to Eqs.~\eqref{omega} and \eqref{sigmaOmega}, the spectrum of the current fluctuations in superconductors is dominated by the $1/\mathrm{f}$ noise for  $\omega <\omega^{*}$. This frequency interval can be significantly larger than the interval in which the $1/\mathrm{f}$ noise exceeds the equilibrium Johnson-Nyquist noise in the normal metals.

Equations \eqref{eq:epsilon_p_s}, \eqref{eq:delta_epsilon_omega}   and \eqref{eq:delta_epsilon_alpha_s_estimate} can be obtained from an alternative consideration by evaluating the sensitivity of the superfluid density per unit area,  $N_{s}$,  to a change of state of the TLS.
It was shown in Refs.~\cite{AltsSpivNs, ZyuzSpivNs, ZyuzSpivRev} that the r.m.s. the amplitude of its  mesoscopic fluctuations, of the superfluid density
$\delta N_{s}=N_{s}- \langle N_{s} \rangle $, averaged over   a region of size $\xi$,
scales with the amplitude of universal  conductance fluctuations~\cite{AltSpivSenc,LeeStone} of the sample,
\begin{align}\label{eq:delta_N_s}
\sqrt{\langle (\delta N_{s})^{2}\rangle} \sim & \,  \langle N_{s} \rangle /G \sim N\lambda_{F}  (\Delta/\epsilon_{F}).
\end{align}
Here $N$  is the electron density,  and we used the expression for average superfluid density per unit area  $\langle N_{s}\rangle \sim  N L_{z} \ell / \xi_{0}$.

By changing the electron interference pattern, transitions of an individual TLS  located at $\bm{r}_i$ induce time-dependence of the superfluid density in the spacial region of the order of the coherence length $\xi$ near $\bm{r}_i$. We denote the time-dependent part of the superfluid density averaged over this region by $\delta N_{s}({\bf r}_{i}, t)\equiv \tilde{N}_{s, i}(t)$.   At $r_{s}\sim 1$ the difference in the electron scattering amplitudes in the two different states of a TLS is of the order of $\lambda_{F}^{2}$. Therefore, using Eq.~\eqref{eq:delta_N_s} the r.m.s. value of a random change of the superfluid density caused by a change of state of an individual  TLS may be estimated as~\footnote{The estimate \eqref{tildeN} can be obtained as follows. It was  shown in Refs.~\cite{AltSpivSenc,FengLeeSen} (see also Ref.~\cite{ZyuzSpivRev} for a review) that the number of impurities $n_0$  whose scattering amplitude needs to be changed by $\delta f \sim \lambda_{F}$  in order  to  change the conductance by $\delta G\sim e^{2}/\hbar $, (and consequently
the superfluid density by $\delta N_{s}$)  is of order $n_{0}\sim G$.
The qualitative explanation of this fact is as follows.  The mesoscopic fluctuations of $\delta N_{s}$  originate from the interference of diffusion paths traveling across a region of size $\xi$. Each diffusive path can be viewed as a tube with a cross-section area $\sim \lambda_{F}^{2}$ surrounding a classical diffusive trajectory with a typical length $\xi^{2}/\ell $. Thus, its volume is of the order of $v=\xi^{2} \lambda_{F}^{2}/\ell $. The interference pattern changes completely if each diffusion path contains at least one impurity which changes its scattering amplitude. Thus $n_{0}\sim \frac{V}{v}=G$ , where  $V=\xi^{2}L_{z}$ is the volume of a sample of a lateral size $\xi$.
Since the changes of the superfluid density associated with motions of individual scatterers have random signs, the estimate for the r.m.s. value of $\delta \tilde{N}_{s}$, Eq.~\eqref{tildeN}  is obtained by dividing $\delta N_{s}$ in Eq.~\eqref{eq:delta_N_s} by $\sqrt{n_{0}}$.   }
\begin{equation}\label{tildeN}
\sqrt{( \delta \tilde{N}_{s,i})^{2}}\sim
\frac{\sqrt{(\delta N_{s})^{2}}}{\sqrt{G}} .
\end{equation}
Finally, associating the change of the energy of the TLS with the change of the superfluid energy of a block of size $\xi$, $\delta \epsilon ({\bf p}_{s})= \xi^{2}\delta \tilde{N}_{s} {\bf p}_{s}^{2}/m$,  we reproduce  Eq.~\eqref{eq:deltaepsilon}.

Using Eqs.~\eqref{eq:delta_N_s}, \eqref{tildeN} we can also obtain the spectrum of the  $1/\mathrm{f}$ current noise in Eqs.~\eqref{sigmaOmega}, \eqref{S}, \eqref{1/f_sup} without appealing to FDT. This consideration elucidates the microscopic origin of the current noise in current-carrying superconductors, by tracing them to the temporal fluctuations of the superfluid density, $\delta \tilde{N}_{s}({\bf r}, t)$, which are induced by the TLS  transitions.

On the spacial scales larger than $\xi$,  the superfluid density can be written  as
 \begin{equation}
 N_{s}({\bf r},t) -\langle N_{s} \rangle =\sum_{i} \delta \tilde{N}_{s,i} ( t)  \xi^{2} \delta ({\bf r}-{\bf r}_{i}).
 \end{equation}
 Here   the index $i$ labels individual TLS. Assuming that transitions of different TLS are uncorrelated, we get the spectrum of fluctuations of $\delta \tilde{N}_{s,i} ( t)$ in the form
  \begin{equation}\label{fluctuators}
\int dt \overline{ \delta \tilde{N}_{s,i}(t) \tilde{N}_{s,j}(0) } e^{i\omega t}=  \delta_{ij} \frac{\tau_{i}    \, \langle ( \delta \tilde{N}_{s})^{2} \rangle}{\pi\left[ 1+(\omega \tau_{i})^{2}\right]}.
\end{equation}

The local current density ${\bf J}_{s}({\bf r}, t)=\langle {\bf J}_{s}\rangle + \delta {\bf J}_{s}({\bf r}, t) $ and superfluid momentum ${\bf p}_{s}({\bf r}, t)={\bf \bar{p}}_{s} +\delta {\bf p}_{s}({\bf r}, t)$  can be determined from the continuity equation  $\bm{\nabla} \cdot {\bf J}_{s}=0$, which after directing the average superfluid momentum  in $x$-direction, and linearization with respect to small fluctuations, can be  written as
\begin{equation}\label{continuity}
\langle N_{s}\rangle \,  \bm{\nabla} \cdot \delta {\bf p}_{s}({\bf r}, t)+\bar{ \bf p}_{s}\cdot  \bm{\nabla}  N_{s}({\bf r},t)=0.
\end{equation}
In this form the problem represented by Eqs.~\eqref{fluctuators}, \eqref{continuity} becomes equivalent to the problem of current fluctuations in a current bias normal conductor in which the local
conductivity fluctuates in space and time \cite{continuity}. Solving these equations and averaging the result over the distribution of relaxation time with the distribution function $f(\tau_{i})\sim 1/\tau_{i}$ and over the position of  $i$-th TLS, ${\bf r}_{i}$,  we reproduce Eq.~\eqref{1/f_sup}.

The central assumption we made in the article is that the TLS  are situated inside superconductors with $r_{s}\sim 1$ and that their size is of order interatomic spacing. Therefore the characteristic dipolar moment of the TLS is of order $e\lambda_{F}$, while the electron scattering cross-section difference between two the states of TLS is of order $\lambda_{F}^{2}$. Recently, other pictures of TLS have been proposed, including electronic traps \cite{Koch, Mooij},
and localized states associated with spacial fluctuations of superconductor order parameter in strongly disordered superconductors  \cite{Faoro,Ioffe,Feigelman,Ustinov}. In these cases, either the characteristic size of the TLS is significantly larger than $\lambda_{F}$, or the TLS  are situated in an insulator close to the superconductor.
Enhancements of the microwave absorption rate and the spectrum of equilibrium
fluctuations, which are proportional to $\bar{p}_{s}^{2}/\omega^{2}$,  exist in these cases as well.
However, in these situations, Eq.~\eqref{eq: Pdebye}  will acquire additional small pre-exponential factors.

The enhancement of the linear longitudinal conductivity proportional to $(\omega^{*}/\omega)^{2}$ can be several orders of magnitude larger than unity.
Because of this,  in a wide interval of frequencies, the dissipative properties of current-carrying superconductors are dominated by the coupling of TLS to the superfluid current. In this frequency interval, the results presented above enable us to evaluate the  \emph{ac} conductivity and the power of $1/\mathrm{f}$ current noise up to a factor of order unity. The linear approximation in the external electric field is valid as long as $eE_{\omega}/\omega \ll \bar{p}_{s}$.
In the nonlinear regime, $p_{d} >eE_{\omega}/\omega > \bar{p}_{s}$,  Eq.~\eqref{eq:epsilon_p_s} is still valid, and the enhancement of the nonlinear absorption coefficient is independent of $\bar{p}_{s}$.

Finally, we would like to mention that the results presented above may be relevant to the physics of superconductor-insulator-superconductor junctions. In the presence of voltage on the junction the current exhibits oscillations in time which are accompanied by oscillations of superfluid velocity in superconducting leads
supplying the current to the junction.  Therefore there is a contribution to the total dissipation rate of the system associated with TLS in the bulk of the leads.
\textit{Acknowledgements:} Authors are grateful for helpful conversations
with C. Boettcher, A. Burin, M. Feigelman, L. Faoro, L.  Ioffe, S. Kubatkin, C.  Marcus, J. Pekola, and B. Shklovskii.
The work of  B.S. is supported by the DARPA  grant  GR049687, A.A. was supported by the NSF grant  DMR-2424364, and the work of  T. Liu was supported by NSF Grant No. DMR-2203411.

\bibliography{TLSsupercond_Arxiv_1_23}

\end{document}